\documentclass[aps,preprint]{revtex4}%
\usepackage{amsfonts}
\usepackage{amsmath}
\usepackage{amssymb}
\usepackage{graphicx}%
\begin{document}
\title{{\LARGE On the Creation of Universe out of Nothing}\\ }
\author{Marcelo Samuel Berman$^{1}$ \ and \ Luis Augusto Trevisan$^{2}$ }
\affiliation{$^{1}$Instituto Albert Einstein/ Latinamerica\ - Av. Candido Hartmann, 575 -
\ \# 17 \ \ \ \ \ - \ \ \ msberman@institutoalberteinstein.org}
\affiliation{80730-440 - Curitiba - PR - Brazil}
\affiliation{$^{2}$Universidade Estadual de Ponta Grossa \ \ \ \ \ \ \ \ }
\affiliation{Departamento de Matem\'{a}tica/DEMAT - Ponta Grossa - PR - Brazil}
\affiliation{latrevis@uepg.br}

\begin{abstract}
We explain, using a Classical approach, how the Universe was created out of
nothing, i.e., with no input of initial \ energy \ or \ mass. The inflationary
phase, with exponential expansion, is accounted for, automatically, by our
equation of state for the very early Universe.

\end{abstract}
\keywords{Zero energy Universe; Cosmological constant; Inflation; EoS.}\date{Final Version 26 March, 2010.) - (Presented to IWARA09.}
\maketitle

\begin{center}
{\LARGE On the Creation of Universe out of Nothing}

\bigskip

Marcelo Samuel Berman \ and \ Luis Augusto Trevisan
\end{center}

\section{\bigskip Introduction}

The science of cosmology has progressed up to a point where it is possible to
make a valid picture of the first moment of the Universe. It is widely
accepted that the creation of the Universe should be attributed to Quantum
fluctuations of the vacuum. As a consistent Quantum gravity theory is not yet
available, cosmologists employ General Relativity, or other classical
theories, in order to study the evolution of the Universe after Planck's time
\ \  ( $\ \ t\cong10^{-43}s$ $\ \ )$ $\ \ .$

The present authors will try to offer now a Classical picture covering the
creation of the Universe. The only thing that we need to admit, is that
Einstein%
\'{}%
s field equations yield the average values for the quantities that, in the
Quantum Universe, when \ \  $t<10^{-43}s$ $\ \ \ $, fluctuate
quantum-mechanically around those average values, somehow like the Path
\ Integral theory of Feynman\cite{Fey}admits paths that fluctuate around the
average trajectories given by the Classical theory. Throughout this paper,
when we refer to the Quantum Universe, all quantities should be understood
\ as given by \ the most probable values of those quantities, even if this is
not explicitly stated.

Along a rather similar situation, Kiefer, Polarski and Starobinsky
\cite{Staro} have shown that although according to the inflationary scenario
for the very early Universe, all inhomogeneities in the Universe are of
genuine quantum origin, clearly no specific quantum mechanics properties are
observed when looking at this inhomogeneities and measuring them. It looks
like we can get along without a Quantum gravity theory, and still "explain"
the origin of the Universe.

Berman \cite{Berman2009}, shows how to calculate by means of pseudo-tensors,
the zero-energy of the Universe. His calculations date the year 1981
\cite{Berman1981}. References on zero energy calculations for the Universe,
are given by \cite{Berman2009}. It was also shown by Cooperstock and Israelit,
later, \cite{CooIs} that the total energy of the Robertson-Walker's Universe
is zero, without employing pseudo-tensors. The total energy is the sum of the
positive matter energy plus the negative energy of the gravitational field,
and it is zero. Thus, in the very creation moment there was no supply of
energy to the Universe, because it was not needed (see Feynman\cite{10})

\section{The EoS for the very early Universe}

We suppose that the Universe obeyed Einstein's field equations, and
Robertson-Walker's metric. As the field equations yield an expansion, the most
probable value of $R$ (the scale factor) increases, beginning with zero value.
We may suppose that the Universe obeyed the relation %

\begin{equation}
GM=\gamma R
\end{equation}

\noindent where \ \  $\gamma$ \ \ \ is a constant and \ \  $M$ \ \ \ is the
mass of a large "sphere" of radius \ \ \  $R$. \ \ \ When \ \  $t=0$ \ \ \ ,
\ \ \ \ \  $R=0$ \ \ \ \ and, thus, \ \ \ \  $M=0$ \ \ \ . \  As stated in
next section, when relation (1) is plugged with Planck 's length \ \  $R_{pl}$
\ \ \ \ \ and mass \ \ \  $M_{pl}$ \ \ \ \ , we find \ \ \  $\gamma=1$
$\ \ \ \ .$ \ Note that, when $\ \ \ \ \gamma\approx1$ \ \ \ , we have a
Machian Universe\cite{Wein}. So, our Universe needs no initial mass, at least
if it is Machian.

Let us consider the balance of energy equation \cite{Adler}:%

\begin{equation}
p=-\frac{dM}{dV}%
\end{equation}

\bigskip

Using (1) and (2) Berman and Marinho Jr. \cite{Berman} have shown how the very
early evolution of such a Universe obeyed the equation of state
\begin{equation}
p=-\frac{1}{3}\rho\label{prho}%
\end{equation}

\noindent where \ \ \  $V=\alpha R^{3}$ \ \ \ , \ \ \ \ ($\alpha$=const)
\ \ \ , \ \ \  $p$ \ \ \ \ and \ \ \  $\rho\equiv M/V$ \ \ \ \ \ stand for
cosmic pressure and energy density.

We obtain relation (3) above, by finding, from (1), and (2) successively:

\[
p=-\frac{dM}{dV}=-\frac{dM}{dR}\frac{dR}{dV}=-\frac{1}{3\alpha R^{2}}\frac
{dM}{dR}=-\frac{\gamma}{3\alpha G}R^{-2}%
\]
while%

\[
\rho=\frac{M}{V}=\frac{\gamma R}{G\alpha R^{3}}=\frac{\gamma}{\alpha G}R^{-2}%
\]
and then, by comparing the expressions for \ \  $p$ \ \ \ \ and $\rho,$
\ \ relation (3) is obtained. Berman \cite{Berman2006}\cite{Berman2006a}, have
tied the zero total energy of the Universe to the Machian calculation of
energy density and cosmic pressure done above.

We shall now show that the solution of the Einstein's field equations, in the
\  $\Lambda\neq0$ \ \ case, yields the exponential law of expansion, which in
turn accounts for the \textquotedblleft inflationary\textquotedblright\ phase
devised by Guth \cite{Guth} and others.

\section{ Solution for the Scale-Factor}

Berman and Marinho Jr. \cite{Berman} admitted that, for the very \ early
Universe, the \ Machian hypothesis \ \  $\frac{GM}{R}=1$ \ \ ,  which is
exactly valid, when mass \ \  $M$\ \ \ \ \ and scale \ factor \ \  $R$
\ \ \ \ are\ plugged with \ Planck 's mass and length \cite{ASTRO} , could
point out to an equation of \ state \ for the very early Universe, in the
classical domain, and thus, not be just a \ coincidence , \ only \ valid for
the Planck 's Universe. The resulting\ \ \ equation of state, that \ would
\ be \ valid \ in the classical Universe domain, just \ \ after the Quantum
State phase, was derived as equation (\ref{prho}), along with:%

\begin{equation}
\rho=(\alpha GR^{2})^{-1}.\ \ \label{alfa}%
\end{equation}

\bigskip

\noindent where \ \  $\alpha=const$ \ \ \ , that relates volume to \ \ \
$R^{3}$\ \ \ \ , by means of%

\begin{equation}
V=\alpha R^{3}%
\end{equation}

In (\ref{alfa}), \ \ \  $G$ \ \ \ \ stands for Newton 's gravitational
constant, while, in (\ref{prho}), \ \  $p$ and \ \ \  $\rho$ \ \ \ stand
\ respectively for cosmic pressure and energy density.

We now propose that \ these results be accepted also as \ yielding the \ most
probable values of the quantities they represent, even \ in the Quantum \ \ domain.

In \cite{Berman}, Berman and Marinho Jr. considered the \ \  $\Lambda=0$
\ \ \ case, obtaining, from Einstein 's field equations, and
Robertson-Walker's metric, the relation%

\begin{equation}
R=\sqrt{\left(  \frac{8\pi}{3\alpha}-k\right)  }.t\label{L0sol}%
\end{equation}

\bigskip

where \ \  $k=0,$ $\ \ \pm1$ \ \ \ is the tricurvature.

For the \ \  $\Lambda\neq0$ \ \ \ case, Einstein's equations read, \ for a
perfect \ fluid:

\begin{equation}
\frac{8\pi G}{3}\rho+\frac{\Lambda}{3}=+\frac{k}{R^{2}}+\frac{\left(  \dot
{R}\right)  ^{2}}{R^{2}}\label{8pi}%
\end{equation}

\noindent and also%

\begin{equation}
8\pi Gp-\Lambda=-\frac{2\ddot{R}}{R}-\frac{(\dot{R})^{2}}{R^{2}}-\frac
{k}{R^{2}}\label{gro}%
\end{equation}

\bigskip

By \ assuming equation of state (\ref{prho}), and relation (\ref{alfa}), we
find the following solution%

\begin{equation}
R(t)=e^{\beta t}-e^{-\beta t}\label{rt}%
\end{equation}

\noindent where \ \  $\beta=const$ \ \ \ \ (to be determined) .We can check
that \ \ \  $R(0)=0$ $\ \ .$

On \ plugging back into equations (\ref{8pi}) and (\ref{gro}), we find%

\begin{equation}
\beta=\frac{1}{2}\sqrt{\left(  \frac{8\pi}{3\alpha}-k\right)  }.
\end{equation}

We can also obtain a value for $\Lambda,$%

\begin{equation}
\Lambda=\frac{3}{4}\left(  \frac{8\pi}{3\alpha}-k\right)
\end{equation}

\bigskip

When \ \  $\Lambda$ \ \ \ \ goes to zero, \ (or \  $\beta t$ \ \ \ is small,
in (\ref{rt})) we recover result (\ref{L0sol}). Though we \ worked with \ \
$\gamma=1,$ (see equation 1), similar results are obtained with \ \ \
$\gamma\neq1$ $\ \ \ ,$ \ \ if \ \  $\gamma$ \ \ \ \ \ is still a constant.
Notice that we have a graceful entrance into exponential inflationary phase,
just as we enter the Classical domain. In fact, the negative exponential part
of the scale-factor , can be neglected with the passage of time. For
inflationary scenario, we suggest the updated book by Weinberg \cite{Wein2008}.

\section{Discussion of Results and Conclusion}

\bigskip

Our result for the scale factor, was adopted by Berman \cite{Berman2009a}%
\cite{Berman2009b} in a different way: here, we produce it as a solution of
the field equations with the given equation of state; there, he began to write
the scale-factor equation, and , then , he found the most general equation of
state that could be found for that scale-factor. It turned that the equation
of state (3) is a particular one in the new solution.

As far we know, our result is so good that it reduces to a known result
(relation (\ref{L0sol})) in the limit \ \  $\Lambda=0$ $\ \ ,$ so we can be
confident on it. We have also shown that the equation of state (3) is valid
not only for \ \ \  $\gamma=1$ \ \ \ , as in reference [6], but for any finite
fixed value of \ \ \  $\gamma$ $\ \ \ .$We remember that \ \  $\Lambda$
\ \ \ \ models \ were considered earlier by Berman and Som \cite{Som}, and
others\cite{Wein2008}.

We explained how, in the origin of the Universe, a "Classical", picture, given
by the most probable values for the Quantum, picture, coinciding with their
classical \ functions, does not require input of initial energy nor initial
mass; just, we need valid the Einstein field equations and R.W's metric .
Thus, we have a consistent picture of the creation of the Universe, out of "nothing".

On the other hand we have found an equation of state and a solution for the
scale factor in the \ \  $\Lambda$ $\neq0$ \ \ \ \ case that implies the
inflationary phase, automatically, for the very early Universe. 

\bigskip

A first version of this paper was originally published in 2001 (Berman and
Trevisan, 2001)  \cite{BermanTrevisan2001}.

\section{\noindent\textbf{Acknowledgments}}

M.S.B thanks support of Geni, Albert and Paula; and the careful typing by
Marcelo Fermann Guimar\~{a}es.

\end{document}